# Simultaneous deep tunneling and classical hopping for hydrogen diffusion on metals

Wei Fang,[1] Jeremy O. Richardson,[2, *] Ji Chen,[3] Xin-Zheng Li,[4, †] and Angelos Michaelides[3, ‡]

[1] *Thomas Young Centre, London Centre for Nanotechnology and Department of Chemistry, University College London, London WC1E 6BT, UK*

[2] *Laboratory of Physical Chemistry, ETH Zurich, CH-8093 Zurich, Switzerland*

[3] *Thomas Young Centre, London Centre for Nanotechnology and Department of Physics and Astronomy, University College London, London WC1E 6BT, UK*

[4] *School of Physics and the Collaborative Innovation Center of Quantum Matters, Peking University, Beijing 100871, P. R. China*

(Dated: September 8, 2017)

## Abstract

Hydrogen diffusion on metals exhibits rich quantum behavior, which is not yet fully understood. Using simulations, we show that many hydrogen diffusion barriers can be categorized into those with "parabolic-tops" and those with "broad-tops". With parabolic-top barriers, hydrogen diffusion evolves gradually from classical hopping to shallow tunneling to deep tunneling as the temperature ($T$) decreases, and noticeable quantum effects persist at moderate $T$. In contrast, with broad-top barriers quantum effects become important only at low $T$ and the classical to quantum transition is sharp, at which classical hopping and deep tunneling both occur. This coexistence indicates that more than one mechanism contributes to the quantum reaction rate. The conventional definition of the classical to quantum crossover $T$ is invalid for the broad-tops, and we give a new definition. Extending this we propose a model to predict the transition $T$ for broad-top diffusion, providing a general guide for theory and experiment.

Hydrogen (H) diffusion on surfaces is fundamental in disciplines such as surface science, astrophysics, and catalysis [1–9]. Due to the light mass of hydrogen, the process can exhibit significant quantum nuclear effects such as tunneling and isotope effects. The development of surface sensitive techniques means that it is possible to characterize these diffusion processes with high-resolution, and to understand the quantum nature of hydrogen diffusion [10–13]. Various techniques have been applied, including field emission microscopy (FEM) [10], laser optical diffraction (LOD) [11], scanning tunneling microscopy (STM) [12] and helium spin echo (HeSE) [13]. Generally such measurements have been performed on metals because these afford the opportunity of examining diffusion on ultra-clean and atomically flat surfaces, which give the greatest opportunity of revealing fundamental insight of broad relevance.

Several impressive experimental studies have been performed for H and at times deuterium (D) on substrates such as Ni [10, 11, 14], Cu [12, 15], Pt [16, 17] and Ru [18]. Diffusion rates have been measured and upon examining how the rates vary with temperature ($T$), qualitatively different behavior has been seen upon moving from one substrate to another. Relatively straightforward behavior is seen on e.g. Pt(111) where according to HeSE measurements, the rate drops as $T$ is lowered [17]. On Ru(0001), a gradual transition from Arrhenius behavior to a $T$ independent regime has been reported [18]. However on Ni(100) [10] and Cu(100) [12], diffusion rates suddenly become $T$-independent below a certain $T$, indicating a sharp classical to quantum transition. Computational techniques provide complementary insight [19, 20], and previous studies have helped to explain the behavior observed in specific experiments [18, 21–29]. For example the sharp transition on Ni(100) was attributed to the particular shape of the diffusion barrier [21–24]. However previous studies have generally focused on specific surfaces and often force fields have been used. A thorough *ab initio* comparison of hydrogen diffusion on different surfaces (including ones yet to be measured experimentally) is lacking. Moreover, a general physical understanding of surface diffusion and quantum to classical transitions has yet to be obtained.

In this work, we study hydrogen diffusion on metal surfaces with density functional theory (DFT). A key qualitative finding of this study is that many of these processes can be categorized as having barriers with conventional "parabolic-tops" or unconventional "broad-tops". Of the substrates considered, parabolic-top diffusion barriers exist on Cu(111), Ni(111), and Pd(111). When $T$ decreases, the dominant diffusion mechanism evolves gradually from classical over-the-barrier hopping, through shallow tunneling through half of the barrier, to deep tunneling at the barrier bottom. Shallow tunneling enables noticeable isotope effects at moderate $T$ (∼200 K). Broad-top diffusion barriers exist on Ni(100), Cu(100), Ni(110), and Pd(110). For these barriers quantum effects are important only at low $T$ and the classical to quantum transition is sharp, during which classical hopping and deep tunneling coexist. In contrast to the parabolic-top barriers, a rapid onset of isotope effects is predicted for the broad-top barriers. Theoretically, the coexistence indicates that multiple quantum transition states (TSs) can contribute to the reaction rate, providing challenges to quantum rate theories. Using the general insights obtained, we develop a simple model to predict the classical to quantum transition $T$ for broad-top barriers and discuss it within the context of previous experiments and simulations.

Our DFT calculations were carried out using the VASP [30] code with the Perdew-Burke-Ernzerhof (PBE) exchange-correlation functional [31]. We use the nudged elastic band (NEB) method [32] to obtain minimum energy pathways (MEPs) for diffusion (with the substrate atoms and H free to relax in all directions). The MEPs are then used as one dimensional (1D) potential barriers upon which the exact transmission probabilities $P$ for incoherent tunneling are calculated by solving the Schrödinger equation [33]. There are various ways to simulate surface diffusion [19–22, 28, 34], e.g. assuming discrete levels [28, 34], however for a qualitative understanding of the tunneling processes, the initial state distribution is not crucial and depends on experimental conditions. Hence here we use a continuous distribution for simplicity. Additionally, three Feynman path-integral (PI) based approximate theories were tested [35], namely ring-polymer molecular dynamics (RPMD) [36–42], thermalized microcanonical instanton (TMI) [43] and the conventional instanton method [44–48]. The 1D TMI rate is:

$$k_{\text{TMI}} Z_{\text{r}} = \frac{1}{2\pi\hbar} \int_0^\infty e^{-W(E)/\hbar} e^{-\beta E} dE, \quad (1)$$

where $W(E) = S(\tau) - \tau E$ is the abbreviated action, $S$ the Euclidean action of the instanton, $\tau$ its imaginary time, $E = \frac{\partial S}{\partial \tau}$ its energy, $\beta = \frac{1}{k_B T}$, $k_B$ the Boltzmann constant, and the scattering partition function is used for $Z_{\text{r}}$. $W(E) = 0$ is used for $E$ larger than the barrier height. The conventional instanton rate, which has been applied to gas phase reactions [4, 49–53], is the steepest-descent (SD) approximation to the integral in the TMI rate (Eq. 1) [48], hence we refer to it as the SDI. In 1D, the transmission probability $e^{-W/\hbar}$ in Eq. 1 is equivalent to the WKB approximation [54]. Further details of the calculations, validation work, tests with other DFT functionals, and convergence tests are provided in the supporting information (SI) [55].

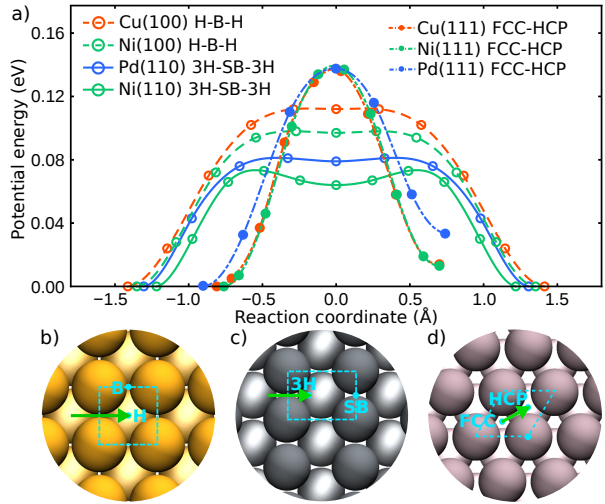

FIG. 1. a) Energy barrier of the H diffusion paths, obtained from NEB calculations using DFT, for several transition metal surfaces. The filled symbols show data for the conventional barriers that are parabolic near the top, and the open symbols are the data points for broad-top barriers. b) Top view of the (100) surface. c) Top view of the (110) surface. d) Top view of the (111) surface. Green arrows show the diffusion paths.

DFT calculations show that the diffusion barriers obtained have two different shapes (Fig. 1(a)). Those between the three-fold hollow sites on Cu(111), Ni(111), Pd(111) (Fig. 1(a)), and Ru(0001) [18] have a conventional parabolic shape near the top. Those on the (100) surface (Fig. 1(c)) of Cu and Ni, and along a path on the (110) surface (Fig. 1(b)) of Pd and Ni, however, are considerably broader. We label these barriers as parabolic-top and broad-top respectively. More examples of both kinds of barrier can be found elsewhere [56]. We see that broad-top barriers can occur when the adsorption sites are relatively far apart (>2.5 Å). Such barrier profiles are possible because unlike covalent bond breaking, a strong bond between H and the continuum of metal states is maintained throughout the diffusion pathway. TSs for typical proton-transfer reaction barriers have an imaginary frequency along the reaction coordinate of ca. $10^3$ cm$^{-1}$. However, for the broad-top barriers discussed here, the TSs have almost zero imaginary frequency. In the case of Ni(110), the barrier top is even a shallow minimum.

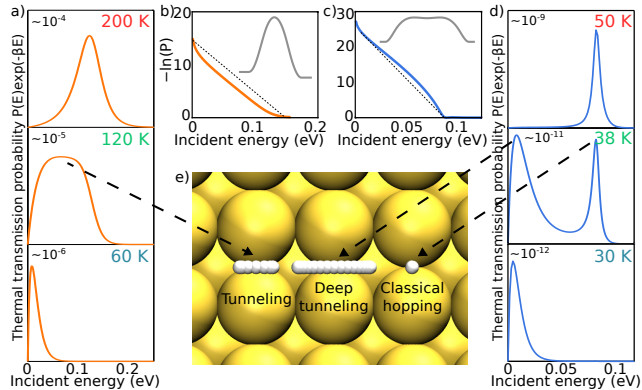

FIG. 2. Temperature dependence of H diffusion on metals for a conventional barrier that is parabolic near the top (Ni(111), left) and one that has a broad-top (3H-SB-3H path Pd(110), right). a) and d), the exact thermal transmission probability $P(E)e^{-\beta E}$ (dimensionless) plotted against the incident energy $E$ for the conventional barrier and the broad-top barrier respectively. b) and c), the transmission action (in units of $\hbar$) defined as $-\ln(P)$, as a function of the incident energy $E$ for the parabolic-top barrier and the broad-top barrier. e) Illustration of the peaks in the thermal transmission probability using tunneling paths represented by the Feynman PI.

We analyze the H diffusion mechanism across two example barriers: diffusion from a pseudo 3 fold hollow site over a short bridge site to another pseudo 3 fold hollow site (3H-SB-3H) (Fig. 1 (c)) on Pd(110) is chosen as the example of a broad-top barrier. We compare it with the parabolic-top barrier found on Ni(111). The results of this analysis are shown in Fig. 2(a) and (d), where we plot the thermal transmission probability $P(E)e^{-\beta E}$ as a function of the incident energy $E$ at $T$s above, during, and below the classical to quantum transitions. For Ni(111), the transmission mechanism changes gradually from being dominated by classical hopping, through shallow quantum tunneling, to deep quantum tunneling as $T$ decreases (Fig. 2(a)). At 200 K when classical hopping is dominant, the transmission probability curve has a tail at low incident energy, meaning that shallow tunneling is also significant. For H diffusion across the 3H-SB-3H broad-top barrier on Pd(110), a different transition behavior from classical hopping to deep tunneling is observed. At 50 K or above, H can only classically hop over the barrier, as reflected by the negligible tail of the thermal transmission probability on the low energy side. At lower $T$ (30 K or below), only deep quantum tunneling is allowed. However around an intermediate transition $T$ (38 K, middle of Fig. 2(d)) the thermal transmission probability curve has *two* maxima, meaning that H can deep tunnel through or classically hop over the barrier with similar probability.

To understand the origin of the anomalous tunneling, we compare the transmission action, defined as $-\hbar\ln(P)$ for the two examples. The larger the transmission action, the more difficult it is for H to tunnel through the barrier at a given incident energy. The broad-top barrier has a convex shaped action when plotted against incident energy (Fig. 2(c)), implying that the broad barrier top hinders shallow tunneling. With a parabolic-top barrier, the action versus incident energy function is concave (Fig. 2(b)), indicating that shallow tunneling is favorable. It is the qualitatively different shapes of the transmission action curves for the two classes of barriers that leads to such different tunneling behaviors. Moreover, this distinction between the two classes of barrier enables us to define broad-top barriers precisely as those barriers for which the transmission action versus energy curve is convex. This definition is valid for all barriers considered in this study (Fig. S2), and classical hopping and deep tunneling channels coexist near the classical to quantum transition $T$ when a barrier has a convex transmission action.

The coexistence of classical hopping and deep tunneling on the broad-top barrier indicates that in contrast to the classical transition state theory picture, multiple quantum TSs can be important. To explore how well quantum rate theories describe this behavior, we analyzed a series of 1D barriers (Eq. S3) constructed by varying the potential from a cosine (parabolic-top) shape to a broad-top one. We calculated rates using three PI-based methods (Fig. 3 (a) and (b)), and compared them to the exact rate. On the parabolic-top barrier, all three theories perform well, agreeing with the exact rate within a factor of 3 (Fig. 3(a)). When the barrier top is broad, TMI and RPMD rates agree with the exact rate within a factor of 2, except for the lowest $T$ (30 K), where RPMD underestimates the rate slightly (Fig. 3(b)). However, the SDI underestimates the rate by a factor of 3-10 in the 30-45 K range (Fig. 3(b)) and with an even wider top, the SDI underestimates the rate by 2-3 orders of magnitude. This is because the SD integral over $E$ breaks down when the instanton is close to $E = 0$, and if multiple quantum TSs contribute. This is the case for broad-top barriers with convex $W$ action, the instanton exists either very close to $E = 0$ or collapsed at the top (Fig. 3(c)). (Other periodic orbits do exist but unlike normal instantons [46] these are second-order saddles of the ring-polymer potential.) Hence the SDI does not capture the coexistence of classical hopping and deep tunneling, and fails to accurately describe the rate at low $T$ (Fig. 3(d)). The TMI solves these problems by avoiding the SD integral over $E$ and instead uses several microcanonical instantons for the rate, which seems to be a promising method for treating complex tunneling problems.

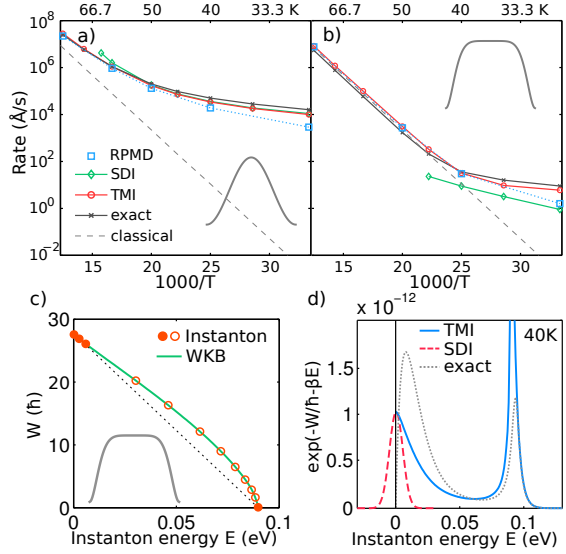

FIG. 3. Performance of different rate theories on 1D barriers (for H) with different shapes. a) Rates on a parabolic-top barrier (inset). b) Rates on a broad-top barrier (inset). The legend is the same as in a). c) The $W$ action plotted against energy $E$ for the broad-top barrier in b). The closed (open) circles show first (second) order saddle instantons obtained through ring-polymer instanton searches. d) The thermal transmission probability at 40 K plotted against energy $E$ for the barrier shown in c) obtained using the TMI theory (blue line) and using the SDI theory (red dashed line).

A key experimental quantity for H diffusion is the classical to quantum transition temperature. Its theory counterpart is the crossover $T$ ($T_c = \frac{\hbar \omega_b}{2\pi k_B}$), defined using the imaginary frequency $\omega_b$ at the barrier top. However $T_c$ is ill-defined for broad-top barriers. Here we define an alternative transition $T$ using the $W$ action in Eq. 1:

$$T_W = \frac{\hbar}{k_B} \frac{E_a}{W_0}. \qquad (2)$$

$E_a$ is the classical activation energy and $W_0 = \oint \sqrt{2mV(x)} dx$, $V(x)$ is the potential. $T_W$ is defined in this way such that $-\frac{1}{k_B T_W}$ is the slope of the dashed line in Fig. 3(c). This means that at $T_W$, the classical hopping (incident energy $E = E_a$) and deep tunneling ($E = 0^+$) have equal contribution to the diffusion rate, hence it is the transition $T$ to deep tunneling

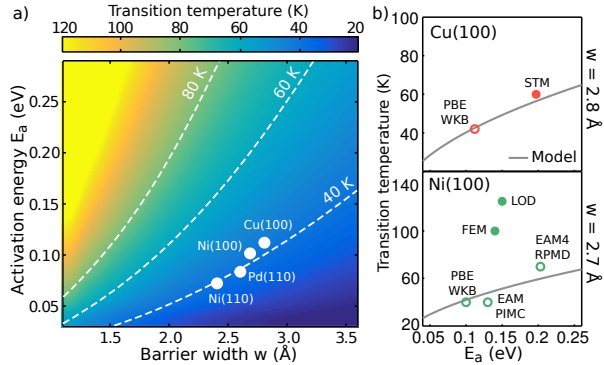

FIG. 4. a) Transition temperature to deep tunneling ($T_W$) predicted by the model over a range of barrier parameters. The 4 broad-top barriers calculated with DFT in Fig. 1 are marked. b) Comparison of the transition $T$ predicted by the model with previous experiments (filled symbols) and theoretical studies (open) on Cu and Ni. Previous results are taken from: STM [12], FEM [10], LOD [11], PI Monte Carlo (PIMC) with an EAM potential [21], and RPMD with an EAM4 potential [24]. The "PBE WKB" points are from this work (Fig. 1).

for broad-top barriers. For barriers with activation energy $E_a$ and width $w$, Eq. 2 becomes:

$$T_W = \frac{\hbar}{n_s k_B}\sqrt{\frac{E_a}{2mw^2}}, \ 0 < n_s < 2, \qquad (3)$$

$n_s$ is a barrier shape factor and $m$ is the mass of hydrogen, and on parabolic-top barriers, $T_W$ and $T_c$ are closely related (derivations see SI). For the model potentials in Eq. S3, when the action curve becomes convex, $n_s \gtrsim 1.5$. Therefore we used $n_s = 1.5$ and plotted Eq. 3 over a range of $E_a$ and $w$ (Fig. 4(a)). This model allows one to estimate the transition $T$ for broad-top barriers without performing rate calculations and is based only on quantities measurable in experiments: the activation energy $E_a$ and the barrier width $w$.

We now use the model and insight obtained to explain and possibly predict the transition $T$ to deep tunneling for H diffusion on several surfaces (Fig. 4(b)). On Cu(100), STM has revealed a classical barrier for H hopping of ∼0.2 eV and a sharp classical to quantum transition at 60 K [12]. Using the experimental $E_a$ our model predicts a transition $T$ of 56 K, in excellent agreement with experiments. Using the DFT barrier, along with the WKB approximation, the transition temperature predicted is ∼40 K. This is still in reasonable agreement with experiments and also consistent with the simple model. On Ni(100) the experimental transition temperatures are in the 100-125 K regime [10, 11]. Previous calculations using force fields have however led to predictions in the 40-70 K regime [21, 23–26].

Both our DFT results and the model yield transition $T$s that are consistent with the previous simulations. Indeed even using the experimental $E_a$ reported (∼0.15 eV) our model predicts a transition $T$ of 50 K. It therefore doesn't seem unreasonable to suggest that an experimental reexamination of H diffusion on Ni(100) could be worthwhile. On Pt(111) we show in the SI that our calculations are consistent with the HeSE experiments [17].

Isotope effects are another important aspect of quantum diffusion and a key experimental signature of tunneling. We have examined the H/D isotope effect on a broad-top barrier and a parabolic-top barrier as a function of $T$. On the parabolic-top barrier, the H/D isotope effect appears at moderate $T$ (100 K) and the rate ratio $k_H/k_D$ increases gradually over a wide temperature range (down to 25 K). In contrast for the broad-top barrier, no H/D isotope effect appears until $T_W$ (∼40 K) is reached, then $k_H/k_D$ increases sharply within a narrow temperature window (40-25 K). These general observations are consistent with previous studies on Ni(100) [23–25], and the qualitative difference between the two types of barriers holds for a broad range of barrier heights and widths. The sharp onset of isotope effects could therefore serve as an experimental signature of diffusion on a surface with broad-top barriers.

To conclude, insight into the quantum nature of hydrogen diffusion on metals has been obtained. A clear qualitative distinction between barriers with broad-tops and conventional parabolic-tops has been identified. For the broad-top barriers we observed a regime at which both classical hopping and deep tunneling are favored. Despite the long history of theories for general tunneling phenomena [54, 57], we are not aware of discussion of such behavior. It remains to be seen if similar behavior for hydrogen diffusion and proton transfer will be seen in other environments. Treating more complex systems will likely require a full multidimensional description of the process. The unique behavior observed here has led to a series of general implications, including a requirement for a multi-TS theoretical treatment, a new definition of the classical to quantum transition $T$ ($T_W$) and a sudden emergence of strong isotope effects around $T_W$.

J.C. and A.M. are supported by the European Research Council under the European Union's Seventh Framework Programme (FP/2007-2013) / ERC Grant Agreement number 616121 (HeteroIce project). A.M. is also supported by the Royal Society through a Royal Society Wolfson Research Merit Award. X.Z.L. is supported by MOST 2013CB934600, the National Key R&D Program under Grant No.2016YFA0300901, and the National Science

Foundation of China under Grant Nos 11422431 & 11634001. We are grateful for computing resources provided by the London Centre for Nanotechnology, Research-Computing at University College London, and the UK's HEC Materials Chemistry Consortium, funded by EPSRC (EP/L000202) for access to the ARCHER UK National Supercomputing Service.

---

* jeremy.richardson@phys.chem.ethz.ch

† xzli@pku.edu.cn

‡ angelos.michaelides@ucl.ac.uk

# Supporting Information: Simultaneous deep tunneling and classical hopping for hydrogen diffusion on metals

Wei Fang,[1] Jeremy O. Richardson,[2, *] Ji Chen,[3] Xin-Zheng Li,[4, †] and Angelos Michaelides[3, ‡]

[1]*Thomas Young Centre, London Centre for Nanotechnology and Department of Chemistry, University College London, London WC1E 6BT, UK*

[2]*Laboratory of Physical Chemistry, ETH Zurich, CH-8093 Zurich, Switzerland*

[3]*Thomas Young Centre, London Centre for Nanotechnology and Department of Physics and Astronomy, University College London, London WC1E 6BT, UK*

[4]*School of Physics and the Collaborative Innovation Center of Quantum Matters, Peking University, Beijing 100871, P. R. China*

**OUTLINE**

Here we provide more computational details of the simulations reported in the manuscript and additional discussions in support of the conclusions reached. The tests on the computational setup and validity for our density functional theory (DFT) diffusion barriers are given in section S.I. The transmission action of all the barriers calculated and the Eckart barrier are discussed in section S.II. In section S.III we provide details about the model potentials we used for comparing quantum rate theories, and show further details and results of our RPMD simulations. We discussion the effects of dimensionality in section S.IV. In section S.V we provide more details for the new definition of classical to quantum transition temperature. In section S.VI isotope effects are discussed. In section S.VII we show the discussions on the existing experiments on Pt(111).

**S.I. COMPUTATIONAL SETUP FOR THE DFT CALCULATIONS**

For the transition metals we have considered, DFT correctly predicts the lattice constants and the H adsorption site on each surface examined. The vibrational frequencies computed are also in good agreement with experiments available. For example on Pd(110), electron-energy-loss spectroscopy (EELS) measured two clear peaks at 96-100 and 121-122 meV for hydrogen (and $\sim 1/\sqrt{2}$ of the two values for deuterium) [1]. Our frequency calculations show that they agree spot on with the outer plane vibration at the long bridge site (98 meV) and the pseudo-three-fold hollow site (120 meV). Our calculations on Ni surfaces reproduces experimental adsorption energies reasonably well [2, 3] and agreed very well with previous DFT studies [4].

To obtain reliable H diffusion barriers, the precise DFT setup for each system was determined after a careful set of tests. We examined the convergence of the adsorption energy:

$$E_\text{ad} = E_\text{H+surf} - E_\text{surf} - 1/2 E_\text{H}_2, \tag{S1}$$

with respect to the plane wave energy cutoff, K-points, and number of layers in our DFT calculations on the high symmetry adsorption sites on the various surfaces discussed in the manuscript. We also examined the convergence of the relative energy difference between the high symmetry sites on the surfaces. The top two layers of surface atoms are flexible in our calculations.

| metal | En-cut (eV) | K-points | unit cell | N layers | $E_{ad}$ 3H | $E_{ad}$ SB | $\Delta E$ (eV) |
|---|---|---|---|---|---|---|---|
| Pd | **350** | **4×4×1** | **2×3** | **7** | -0.525 | -0.445 | 0.080 |
| | 350 | 4×4×1 | 2×3 | 8 | -0.512 | -0.410 | 0.102 |
| | 350 | 6×6×1 | 2×3 | 8 | -0.503 | -0.405 | 0.098 |
| | 350 | 4×4×1 | 2×3 | 10 | -0.523 | -0.435 | 0.088 |
| Ni | **400** | **4×4×1** | **2×3** | **8** | -0.432 | -0.368 | 0.064 |
| | 500 | 4×4×1 | 2×3 | 8 | -0.434 | -0.370 | 0.064 |
| | 400 | 6×6×1 | 2×3 | 8 | -0.429 | -0.364 | 0.065 |
| | 400 | 4×4×1 | 2×3 | 10 | -0.442 | -0.376 | 0.066 |

TABLE S1. Convergence of the plane wave energy cutoff, K-points, and number of layers for the DFT calculations on the (110) surface of Pd and Ni. The $E_{ad}$ (units in eV) of the pseudo-3-fold hollow site (3H) and the short bridge site (SB) are compared, and $\Delta E$ is the energy difference between the two sites. The parameters used in the NEB calculations are given in bold.

| metal | En-cut (eV) | K-points | unit cell | N layers | $E_{ad}$ H | $E_{ad}$ B | $\Delta E$ (eV) |
|---|---|---|---|---|---|---|---|
| Ni | **400** | **5×5×1** | **2×2** | **8** | -0.523 | -0.426 | 0.097 |
| | 500 | 8×8×1 | 2×2 | 10 | -0.510 | -0.405 | 0.105 |
| Cu | **400** | **7×7×1** | **2×2** | **8** | -0.177 | -0.065 | 0.112 |
| | 500 | 9×9×1 | 2×2 | 10 | -0.138 | -0.048 | 0.090 |

TABLE S2. Convergence of the plane wave energy cutoff, K-points, and number of layers for the DFT calculations on the (100) surface of Ni and Cu. The $E_{ad}$ of the hollow site (H) and the bridge site (B) are compared, and $\Delta E$ is the energy difference between the two sites. The parameters used in the NEB calculations are given in bold.

The test results are given in Table S1 ((110) surface), Table S2 ((100) surface), and Table S3 ((111) surface). We used a tight force convergence criterion (0.002 eV/Å) in our geometry optimisation and nudged elastic band (NEB) calculations [5]. The parameters used for the NEB calculations in the manuscript are given in bold, which converge the relative energies of the adsorption sites to within 20 meV.

We also tested the impact of different exchange-correlation functionals on the diffusion barriers. To this end, we have performed NEB calculations with the PBE-D3 and

| metal | En-cut (eV) | K-points | unit cell | N layers | $E_{ad}$ FCC | $E_{ad}$ HCP | $\Delta E$ (eV) |
|---|---|---|---|---|---|---|---|
| Ni | **400** | **5×5×1** | **2×2** | **8** | -0.581 | -0.568 | 0.013 |
|  | 500 | 7×7×1 | 2×2 | 10 | -0.552 | -0.542 | 0.010 |
| Cu | **400** | **7×7×1** | **2×2** | **8** | -0.189 | -0.175 | 0.014 |
|  | 500 | 9×9×1 | 2×2 | 10 | -0.207 | -0.192 | 0.015 |
| Pd | 350 | 4×4×1 | 2×2 | 7 | -0.567 | -0.525 | 0.042 |
|  | **350** | **6×6×1** | **2×2** | **8** | -0.583 | -0.550 | 0.033 |
| Pt | **400** | **5×5×1** | **2×2** | **8** | -0.485 | -0.445 | 0.040 |
|  | 400 | 9×9×1 | 2×2 | 8 | -0.501 | -0.462 | 0.039 |
|  | 500 | 9×9×1 | 2×2 | 10 | -0.487 | -0.450 | 0.037 |

TABLE S3. Convergence of the plane wave energy cutoff, K-points, and number of layers for the DFT calculations on the (111) surface of Ni, Cu and Pt. The $E_{ad}$ of the FCC site and HCP site are compared, and $\Delta E$ is the energy difference between the two sites. The parameters used in the NEB calculations are given in bold.

optB88-vdW functional for both a broad-top example (Ni(100)) and a parabolic-top example (Pd(111)). These two functionals incorporate dispersion interactions, which are not captured with the PBE functional that was used throughout this study. The results are shown in Fig. S1(a). Overall we have not found any example where a change in functional led to a qualitative change in the nature of the diffusion profile. Different functionals, as expected, yield slightly different energy barriers but in all cases the broad top barriers remain broad topped and the parabolic barriers remain parabolic.

We also considered the zero point energy (ZPE) of the H diffusion barriers. Two examples, Cu(100) and Ni(100), are shown in Fig. S1(b), one can see that the shape of the barriers changed little. ZPE increased the barrier height, this is because the transition states have more higher-frequency vibrational modes than the initial state.

### S.II. TRANSMISSION ACTION OF THE DFT BARRIERS

The exact transmission probability $P(E)$ for H crossing a 1D barrier as a function of the incident energy $E$ can be calculated by solving the time-independent Schrödinger equa-tion numerically using the Numerov method. $P(E)$ is the key quantity for quantum rate calculations. Integrating the thermal transmission probability $P(E)e^{-\beta E}$ gives the thermal diffusion rate at temperature $T$:

$$kZ_\mathrm{r} = \frac{1}{2\pi\hbar} \int_0^\infty P(E)e^{-\beta E} dE, \qquad (S2)$$

where $\beta = \frac{1}{k_\mathrm{B}T}$, $k_\mathrm{B}$ is the Boltzmann constant and the scattering partition function $Z_\mathrm{r} = \sqrt{\frac{m}{2\pi\beta\hbar^2}}$, $m$ is the mass of H, is used for the 1D barriers. The transmission action (defined as $-\hbar\ln(P)$) of all the barriers in Fig. 1 of the main text are shown in Fig. S2. The broad-top barriers (Ni(100), Ni(110) and Cu(100)) have a convex shape transmission action while the parabolic top barriers (Cu(111) and Pd(111)) have a concave shape transmission action. If the transmission action is exactly linear, at the transition temperature (Eq. S4), classical hopping, shallow tunneling and deep tunneling all contribute to the diffusion rate.

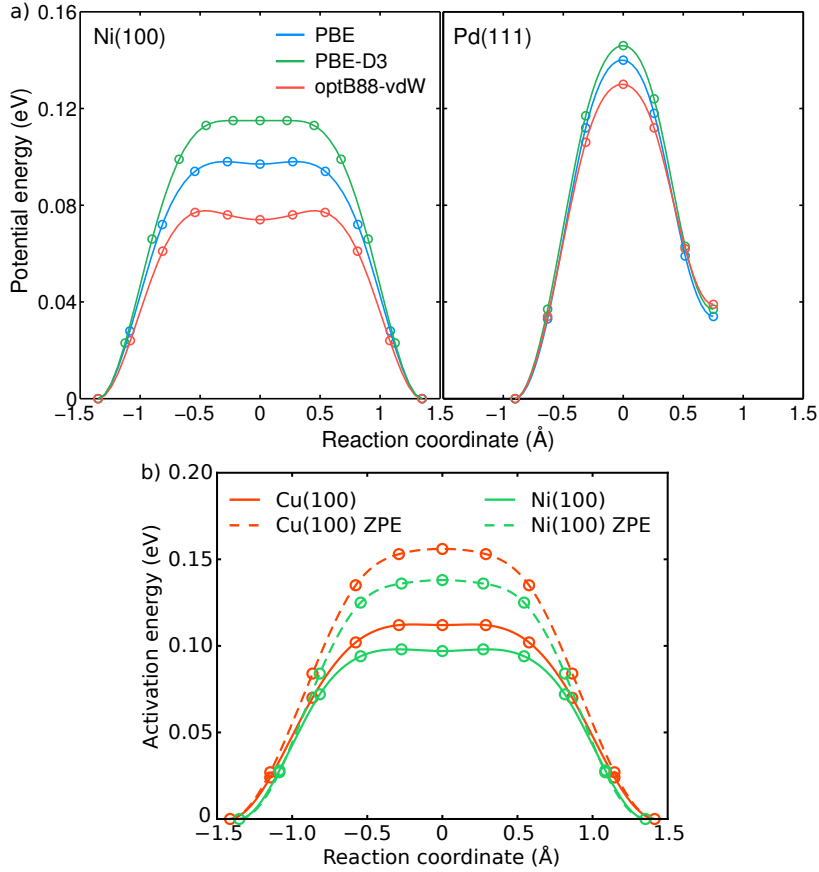

FIG. S1. (a) Comparison of the effect of functionals on the H diffusion barriers. The points are obtained from NEB calculations. (b) Comparison of the impact of ZPE of H on the barrier shape.

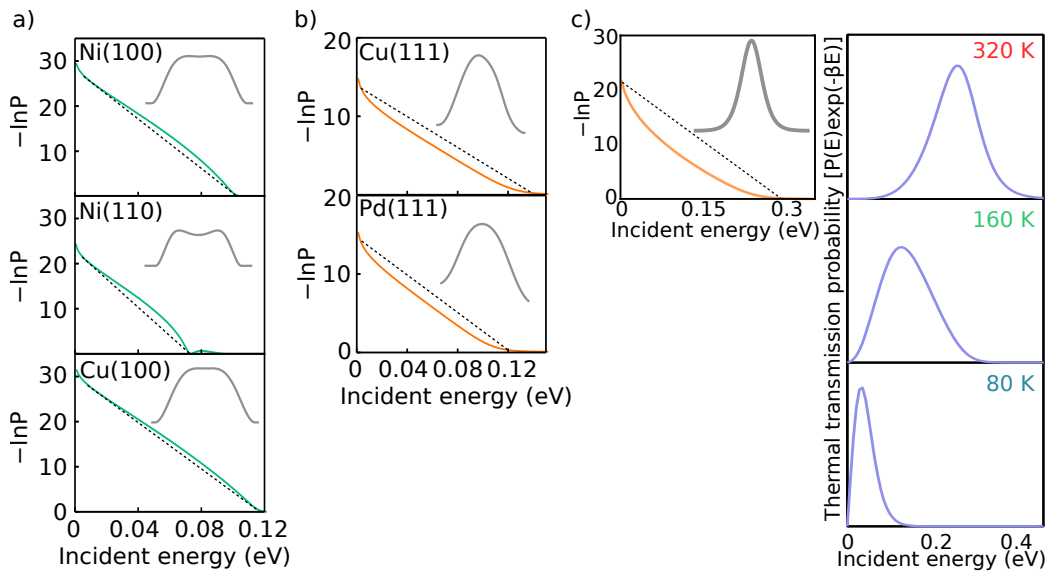

FIG. S2. a) The transmission action (in units of $\hbar$), as a function of the incident energy $E$ for the broad-top barrier examples in Fig. 1 of the main text. They all have a convex shape. b) The exact transmission action of the parabolic-top barrier examples in Fig. 1 of the main text. They all have a concave shape. c) The exact transmission action of an Eckart barrier $V(x) = B \operatorname{sech}^2(x/a)$. The Eckart barrier parameters are the same as in [6], $a = 0.66a_0$, $B = \frac{72\hbar^2}{\pi^2 m a^2}$ ($\sim$0.25 eV), where $a_0$ is the Bohr radius and $m$ is the mass of H, this barrier is $\sim$1 Å wide and the imaginary frequency at the top is $\sim -1000$ cm$^{-1}$. The thermal transmission probability plotted against the incident energy at three different temperatures are shown in the right panels.

## S.III. COMPUTATIONAL DETAILS AND DISCUSSIONS FOR RPMD AND IN-STANTON SIMULATIONS

We have tested the performance of different rate theories on a set of constructed barriers varying from cosine shape to broad-top shape with the functional form:

$$V(x) = E_\mathrm{a} \cos^2 \left[ \left( \frac{2x}{w} \right)^{2r} \frac{\pi}{2} \right] \quad (S3)$$

in which $E_\mathrm{a} = 0.09$ eV, $w = 2.5$ Å, and $2r = 1, 3, 5$. $V = 0$ is used for $|x| > w/2$, as the abbreviated action in the instanton theory [7] and the WKB theory is independent of periodicity [8].

The ring polymer molecular dynamics (RPMD) rate [9], is calculated using the Bennett-Chandler method [10] that separates the rate into a dynamical factor multiplying a free energy term [6, 11]. The free energy barriers were calculated using the potential of mean force method. 60 beads were used for the imaginary time path integral. The same number of beads was used for the instanton calculations, and the rates were converged with respect to the number of beads. A timestep of 0.5 fs was used and we ran a total of 20500 steps at each constrained centroid position, with the first 500 steps discarded for equilibration. With this simulation length, the average forces on the centroid of H have been converged to $10^{-3}$ eV/Å. The free energy barriers obtained are shown in Fig. S3. We see that on the broad-top barriers (2r=3 and 2r=5), the free energy barriers drop suddenly when the temperature decreases from 40 K to 30 K. The dynamical factor is calculated by performing thermostated-RPMD [12] simulations (with the friction factor $\lambda = 0.5$) starting with the ring polymer centroid at the barrier top (dividing surface). [13] The results are given in Table S4. We see that the RPMD dynamical factor is insignificant here as the centroid is a good dividing surface for a symmetric reaction, which is consistent with previous simulations [14].

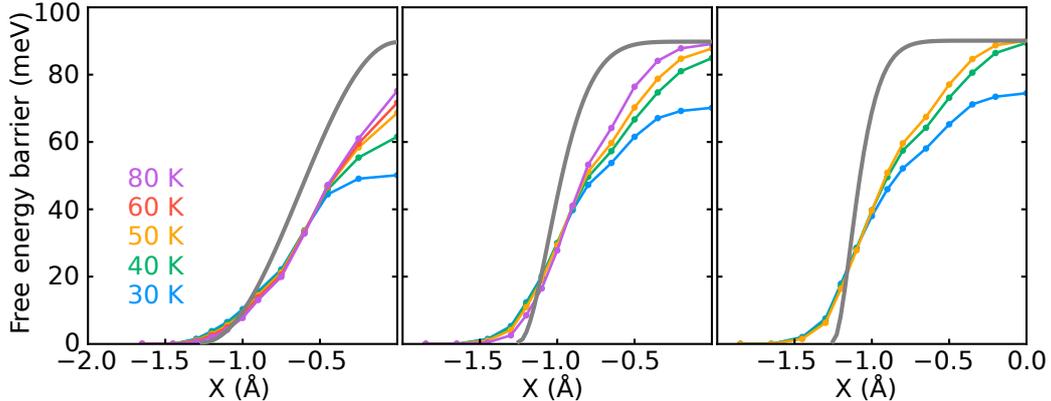

FIG. S3. Free energy barriers for the RPMD rate calculated with the potential of mean force method at different temperatures. The grey lines show the potential energy barriers (Eq. S3), with parameters $2r = 1$ (left), $2r = 3$ (middle) and $2r = 5$ (right).

Fig. S4 shows the comparison of different quantum rate theories on an even broader top barrier (Eq. S3, $2r = 5$) than the one discussed in the main text. The exact rate was computed using Eq. S2. We see more clearly that the conventional steepest-descent instanton (SDI) significantly underestimates the rate (by 2-3 orders of magnitude) and predicts the wrong transition temperature from classical hopping to quantum tunneling, because it fails to describe the physics of the process. SDI is numerically unstable below 35 K as the imaginary frequency tends to 0. SDI is unable to accurately predict the rate at low temperature because the steepest-descent integration over energy is poor. As seen in Fig. 3c in the manuscript, the abbreviated action, $W$ [7, 16], is approximately linear at low energies even though the $dW/dE$ must tend to $-\infty$ in this limit as the periodic orbit starting at $|x| = w/2$ has an infinite period. This implies a very strong contribution from higher derivatives of W which is therefore not well approximated by a second-order Taylor series and the steepest-descent integration. One could imagine generalizing the SDI to include both the low and high energy contributions, but this would not correct the errors in the rate at low $T$. On the other hand, both the RPMD and the recently developed thermalized microcannonical instanton (TMI) agree with the exact results well.

| Temperature (K) | 2r=1 barrier | 2r=3 barrier | 2r=5 barrier |
|---|---|---|---|
| 80 | 1.0 | 1.0 | |
| 60 | 1.0 | | |
| 50 | 1.0 | 1.0 | 1.0 |
| 40 | 0.8 | 1.0 | 1.0 |
| 30 | 0.5 | 0.5 | 0.6 |

TABLE S4. The dynamical factor of the RPMD rate, rounded to one decimal.

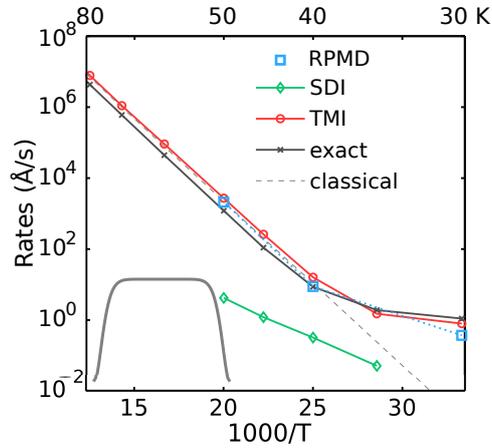

FIG. S4. Comparison of the diffusion rates on a broad-top barrier (Eq. S3) with parameter $2r = 5$, calculated using RPMD, steepest-descent instanton (SDI) [15], and the thermalized microcannonical instanton (TMI) [7] method.

## S.IV. 1D VS MULTI-DIMENSION

Here we discuss the effects of multi-dimensional treatment compared to 1D. First thing we considered is corner-cutting effects, which is interesting and can be important to the instanton path in many situations. For the diffusion paths we studied in this work, due to the fact that the H vibrational frequencies at the transition state perpendicular to the reaction coordinate are very stiff ($> 1000$ cm$^{-1}$) and the instanton has to satisfy symmetry requirements, it very difficult for the instanton path to deviate significantly from the minimal energy pathway. Hence we don't expect corner-cutting effects to be very important here. To show this we performed a full dimensional instanton optimization for the broad-top barrier on Pd(110) and compared the abbreviated action W to a 1D barrier with the same DFT setup (Fig. S5). It is clear that the W action barely changes going from 1D to multi-dimension. Because the W action is well-estimated by the 1D instanton, despite we do not expect our 1D calculations to give quantitative rate predictions, we do expect it to be able to predict the shape of the Arrhenius plot including the crossover temperature.

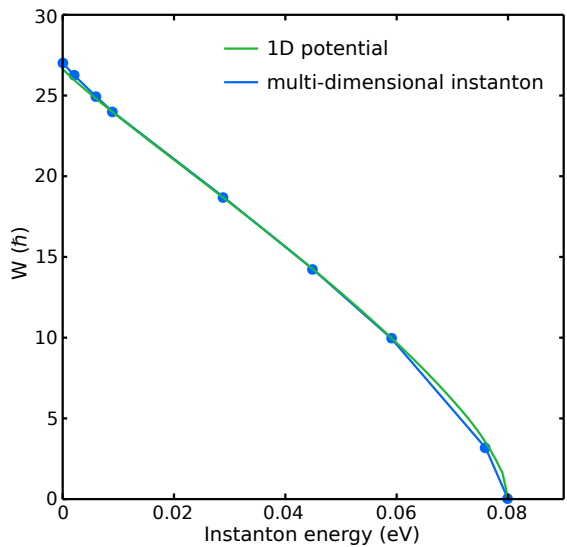

FIG. S5. Abbreviated action W versus instanton energy E for the 1D instanton and multi-dimensional instanton obtained on-the-fly with DFT for the short bridge H diffusion path on Pd(110).

Furthermore, we made comparison of 1D rates with full dimension rates and experimental rates. In our on going work, we have applied full dimensional TMI calculations to Pd(110) surface. We found that the 1D rates can be reasonably good if the reactant and the transition state have similar zero point energies, which is the case for H diffusion over the short bridge path of Pd (110) (they differ by 6 meV). We also calculated 1D rates using a model barrier (Eq. S3, $2r = 2$) with experimental barrier height and width for Cu(100), which reasonably reproduced the STM experimental rates [17]. However we do not expect 1D rates to be always good and certainly do not want to promote using 1D model for accurate rate calculations.

Finally we consider effects of the surface atoms and heat bath. Previous path integral molecular dynamics studies [14, 18] have compared flexible and fixed surfaces for H diffusion on Ni(100) and the rates only change slightly. In addition, we have compared both a flexible and a fixed Cu(100) surface, where the barrier height is 112 meV and 118 meV for the flexible and fixed surface respectively. This also indicates that the movement of surface atom does not have a strong impact on the barrier. We also think including a heat-bath won't change our findings because the barrier top will always remain non-parabolic.

**S.V. DISCUSSIONS ON THE TRANSITION TEMPERATURE FOR BROAD-TOP BARRIERS**

In the manuscript, we have used the abbreviated action $W$ to define an alternative transition temperature for the broad-top barrier:

$$T_W = \frac{\hbar}{k_B} \frac{E_a}{W_0}, \tag{S4}$$

where $W_0 = \oint \sqrt{2mV(x)} dx$ is the abbreviated action at incident energy $E = 0$. Here we show a simple derivation of an easy-to-apply equation of $T_W$ for barriers with barrier height $E_a$ and barrier width $w$:

$$\begin{aligned}
T_W &= \frac{\hbar}{k_B} \frac{E_a}{W_0} \\
&= \frac{\hbar}{k_B} \frac{E_a}{2\int_{-w/2}^{w/2} \sqrt{2mV(x)}dx} \\
&= \frac{\hbar}{k_B} \frac{E_a}{2\sqrt{2mE_a}\frac{w}{2}\int_{-1}^{1}\sqrt{\frac{V(u)}{E_a}}du}, \quad u = \frac{2x}{w} \\
&= \frac{\hbar}{n_s k_B}\sqrt{\frac{E_a}{2mw^2}} \\
n_s &= \int_{-1}^{1}\sqrt{\frac{V(u)}{E_a}}du \in (0, 2)
\end{aligned} \quad (S5)$$

$n_s$ is a barrier shape factor that gives 2 for a square barrier and smaller values as the barrier becomes more curved.

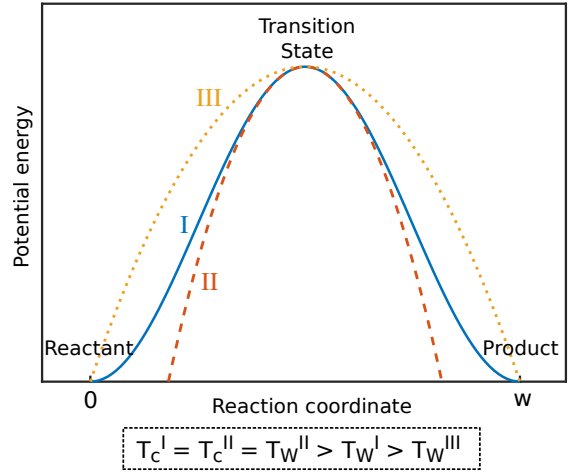

FIG. S6. Relations between $T_c$ and $T_W$ for a parabolic-top barrier. Barrier I is a parabolic-top barrier with width $w$ and activation energy $E_a$. Barrier II is a parabola barrier with the same frequency at the transition state as barrier I. Barrier III is a parabola with width $w$ and activation energy $E_a$. The $T_c$ and $T_W$ on these barriers follow the simple relation: $T_c^{\text{I}} = T_c^{\text{II}} = T_W^{\text{II}} > T_W^{\text{I}} > T_W^{\text{III}}$.

Here we discuss the relation between the conventional crossover temperature $T_c = \frac{\hbar\omega_b}{2\pi k_B}$ and $T_W$ for parabolic-top barriers. On a strict parabola $V(x) = E_a - \frac{1}{2}m\omega_b^2 x^2$, the shape factor $n_s = \frac{\pi}{2}$, and the barrier width $w = \sqrt{\frac{8E_a}{m\omega_b^2}}$. Plugging into Eq. S5, one can find that for a strict parabola, $T_c = T_W$. Hence on parabolic-top barriers, the relation between $T_c$ and

S11

$T_W$ is shown straightforwardly in Fig. S6. In summary, for broad-top barriers, $T_W$ is well-defined and has a clear physical picture while $T_c$ is not; on strict parabolas, $T_c = T_W$; and on parabolic-top barriers, both $T_c$ and $T_W$ are well-defined and closely related ($T_c > T_W$), but $T_c$ has a clearer physical picture than $T_W$. The most suitable transition temperature for all barriers would ideally be given by combining $T_c$ and $T_W$. For example, use the larger value between $T_c^{\rm I}$ and $T_W^{\rm I}$ (or $T_W^{\rm III}$).

**S.VI. ISOTOPE EFFECTS**

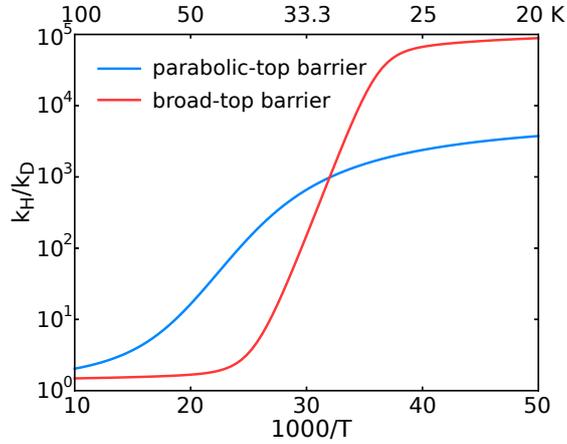

FIG. S7. Comparison of the H/D isotope effect, namely the diffusion rate ratio between H and D, on a broad-top barrier (Eq. S3, $2r = 3$) and on a conventional parabolic-top barrier (Eq. S3, $2r = 1$).

Fig. S7 shows a comparison of the H versus D diffusion rates, reported as $k_{\rm H}/k_{\rm D}$. Significant differences in the rates of H and D diffusion emerge rapidly within a narrow temperature window for broad-top barriers, whereas this happens much more gradually for parabolic-top barriers.

**S.VII. MORE ON EXPERIMENTAL IMPLICATIONS**

Fig. S8 shows our results on Pt(111). On this surface, HeSE [19] measured that the H diffusion rates is still temperature dependent even at 80 K and displays small quantum effects (only 1 order of magnitude higher than the classcial rate at 100 K), while LOD [20] experiment showed transition to temperature independent tunneling at 95 K. We calculated two diffusion pathways on this surface with NEB (Fig. S8(a)). The pathway FCC-Top has a much higher energy barrier than the pathway from FCC-HCP and hence is not expected to contribute much at these temperatures. The pathway FCC-HCP has an energy difference of 40 meV between the FCC and HCP sites and there is a small barrier at the bridge. Only shallow tunneling from FCC to HCP can happen and the rate should be temperature dependent down to very low temperatures when deep tunneling from FCC to another FCC occurs or deep tunneling between the FCC and top site becomes competitive. The rates calculated on the DFT NEB barrier for Pd(111) also agree very well with the HeSE experiment rates (within a factor of 5). In this sense our results support the HeSE measurements.

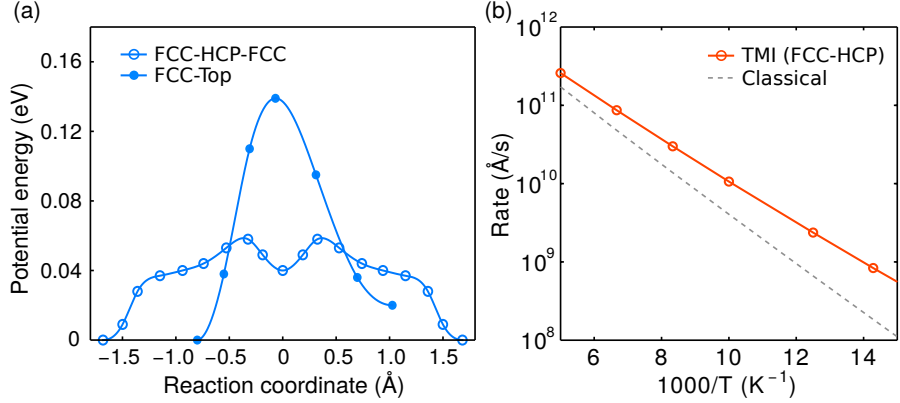

FIG. S8. (a) Potential energy barriers of H diffusion pathways on Pt(111). (b) The diffusion rate estimated using the TMI method for the FCC-HCP pathway in the temperature range 70 - 200 K.

---

* jeremy.richardson@phys.chem.ethz.ch

† xzli@pku.edu.cn

‡ angelos.michaelides@ucl.ac.uk